\documentclass[sigconf]{acmart}
\usepackage{booktabs}
\usepackage{array}
\usepackage{tabularx}
\usepackage{makecell}
\usepackage{amsmath}
\usepackage{amsfonts}
\usepackage{multirow}
\usepackage{xcolor}
\usepackage{algorithm}
\usepackage{algorithmic}
\usepackage{listings}
\usepackage{float}
\usepackage{tikz}
\usetikzlibrary{arrows.meta,calc,positioning}
\AtBeginDocument{%
  }

\setcopyright{none}
\copyrightyear{2027}
\acmYear{2027}
\acmDOI{}
\acmISBN{}
\acmConference[KDD '27]{The 33rd ACM SIGKDD Conference on Knowledge Discovery and Data Mining}{August 1--5, 2027}{San Jose, CA, USA}

\begin{document}

\title{Verify, Don't Trust: Agentic Model Development for Video Discovery Retrieval at Scale}

\author{Hao Fu \quad Baiting Zhu \quad Minglei Chen \quad Yinjie Huang \quad Shuai Ding}
\affiliation{%
  \institution{Meta Platforms, Inc.}
  \city{Menlo Park}
  \state{California}
  \country{USA}}

\makeatletter
\gdef\authors{Hao Fu\and Baiting Zhu\and Minglei Chen\and Yinjie Huang\and Shuai Ding}
\makeatother

\renewcommand{\shortauthors}{Fu et al.}

\begin{abstract}
Large language model (LLM) agents can now propose, implement, and evaluate model changes. Existing autoresearch loops demonstrate this capability through minutes-scale iterations on a self-contained program. Online autoresearch instead spans asynchronous systems, hours-long variants, and weeks-long campaigns that can influence a product. A completed run can still support an invalid conclusion when a code change is a no-op, data windows leak, evaluator semantics drift, or the two arms traverse different serving funnels. We present \textsc{EvoPilot}, a human-gated method for \emph{long-horizon online autoresearch}. Role-specific agents plan and execute each round through a versioned domain skill and typed adapter. Durable records preserve experiments and failures across rounds, while deterministic checks verify the resulting artifacts. Retrieved memories guide agents, but only executable checks enforce recorded lessons.

We study a 37-day campaign for the retrieval system that powers Video Deep Dive (VDD), an online experience for discovering follow-on videos after a user opens a seed video. The campaign covered seven directions and used an hourly refreshed index of hundreds of millions of videos. Earlier manual experiments had not established a benefit from an interaction head. A primitive autoresearch attempt revisited the direction but incorrectly attributed an offline hit-rate decline of 22 percentage points to the head. We then introduced \textsc{EvoPilot}. Its human-gated verification traced the drop to a pre-existing evaluation defect that produced output depths of 3,000 and 600. After repair, a matched comparison measured an offline improvement of 3.20 percentage points. Post-study replay and mutation tests rejected invalid comparisons while admitting valid counterparts. Durable state recovered an interrupted round, and artifact reuse avoided approximately five GPU-hours. Separately, a seven-day randomized online evaluation estimated a 0.66\% relative increase in the VDD slice of Good Search Result Rate for Retention (GSRR).
\end{abstract}



\maketitle

\section{Introduction}
\label{sec:intro}

Large retrieval models are developed through sequences of experiments, not a single training run. Existing agentic-ML systems automate experiment modification and paper generation~\cite{huang2024mlagentbench,lu2024aiscientist}; fixed-compute loops repeatedly edit a self-contained program and accept or revert each minutes-scale run using one scalar metric~\cite{karpathy_nanochat}. We study \emph{long-horizon online autoresearch}, where a candidate crosses independently evolving source, data, training, publishing, serving, and evaluation systems. Variants run for hours and campaigns span weeks. The question is not only whether an agent finds a higher number, but whether durable evidence supports the intended comparison.

Completing an experiment does not by itself make the result valid. Over days or weeks, an intended code change may not take effect, a data partition may be replaced, or an evaluator may change the meaning of its metric. Control and treatment runs can each complete successfully yet still be incomparable if they use different code versions, data snapshots, or candidate depths. Remote workflows may also fail partway or outlive an agent session, so recovery must preserve valid upstream work. Addressing these risks requires durable experiment state and executable checks against produced artifacts, rather than instructions in an agent prompt.

We study these challenges in Video Deep Dive (VDD), an online follow-on discovery experience serving hundreds of millions of users. After a user opens a seed video, VDD retrieves videos for further exploration using the seed, a machine-generated pivot query, and user context. Its model-based retrieval (MBR) system runs on an existing GPU serving platform backed by an index of hundreds of millions of videos refreshed hourly. Our analysis concerns the experiment-control layer above this retrieval system.

We present \textsc{EvoPilot}, a human-gated system that treats an online experiment as a protocol rather than a prompt. A human approves each plan, and isolated workers execute the variants through a versioned domain skill and adapter. Verifiers check artifact-backed comparisons. An analyst updates the hypothesis tree, and a planner turns the evidence into next-round variants that require fresh approval. An event ledger, prior results, experiment narratives, and an incident catalog preserve context across rounds. This memory remains advisory until a lesson becomes an executable check over an artifact or state transition.

We studied a $\sim$37-day VDD campaign across seven research directions. In one case, manual experiments had not established a benefit from interaction-head rescoring. A primitive autoresearch attempt revisited the direction but incorrectly attributed a $22$\,pp offline drop to the head. After EvoPilot was introduced, its human-gated verification found that the control realized $K_{\mathrm{eff}}{=}3000$ and the treatment $K_{\mathrm{eff}}{=}600$, despite both declaring $K{=}600$. A repaired, matched comparison measured a $3.20$\,pp improvement. The selected model later increased VDD GSRR by 0.66\% relative in a randomized online A/B test.

This paper makes three contributions:
\begin{itemize}
    \item \textbf{A protocol for long-horizon online autoresearch.} We distinguish a completed workflow from a valid comparison. The protocol records the shared base, declared treatment, data snapshot, evaluator, metric, and post-conditions needed to interpret an hours-long result days later.
    \item \textbf{A working realization.} Evaluated internally for VDD model development, \textsc{EvoPilot} combines scoped agent roles, a versioned MBR procedural skill, an execution adapter, isolated variants, persistent evidence and memory, fail-closed comparison admission, and a process that turns incidents into guards. Documented outcomes include human-gated rejection of the historical output-depth mismatch, recovery of interrupted work, and artifact reuse that avoids redundant publication.
    \item \textbf{A longitudinal field study.} A 37-day inventory covers 20 approved rounds in seven VDD directions, with a record-complete subset used for operational analysis. Paired replay/mutation tests exercise ten safeguards; a separate randomized A/B test establishes downstream model impact.
\end{itemize}

\section{Related Work}
\label{sec:related}

We first position the method against agentic experimentation and ML lifecycle systems, then describe the retrieval setting in which it is evaluated.

\subsection{Automated and agentic ML development}
Automated machine learning (AutoML) tunes within a fixed search space. More recent systems use LLM agents as research drivers. MLAgentBench evaluates agents that modify and run ML experiments~\cite{huang2024mlagentbench}; fixed-compute loops edit one script and keep or revert by a scalar~\cite{karpathy_nanochat}; and the AI Scientist automates ideas, experiments, and paper writing~\cite{lu2024aiscientist}. Analyses also identify failures in autonomous experimental reasoning and verification~\cite{trehan2026scientists}. EvoPilot targets the complementary hours-to-weeks regime: execution is an asynchronous remote DAG, valid upstream artifacts must survive partial failures, and evidence must persist across sessions and rounds. The objective is experimental continuity and integrity, not merely a shorter edit--train--score cycle.

Textual reflection~\cite{shinn2023reflexion} and procedural skill libraries~\cite{wang2023voyager} show how agents can retain experience across trials. EvoPilot likewise preserves reusable context, but separates an append-only evidence ledger, program memory for prior rounds and experiments, and incident lessons for future behavior. Stored text remains advisory; a remembered instruction becomes enforceable only after it is encoded as an executable guard.

\subsection{Experiment tracking and ML pipeline quality}
MLflow and ModelDB record parameters, artifacts, and lineage for model runs~\cite{zaharia2018mlflow,vartak2016modeldb}; TensorFlow Extended (TFX) and operational-readiness rubrics add pipeline validation and release checks~\cite{baylor2017tfx,breck2017mltestscore}. These systems improve run traceability and operational readiness, but do not by themselves determine whether two completed runs realize an approved experimental contrast.

Recent work uses \emph{harness engineering} to move reliability requirements from prompts into executable controls. Prior systems enforce enterprise QA protocols~\cite{ahn2026contracts}, check structural certificates across agent-framework compilers~\cite{banu2026categorical}, or combine tests, specifications, simulation, shadow traffic, and telemetry~\cite{datadog2026harness}. EvoPilot applies this principle to a relational experiment protocol. Its verifier checks whether control and treatment share the approved base, data, evaluator, and non-treatment artifacts unless a difference is authorized. The admitted object is therefore a comparison across a multi-stage ML workflow, not an individual run. As Table~\ref{tab:admission_objects} shows, two valid, well-tracked runs can still form a rejected comparison.

\begin{table}[t]
\centering
\scriptsize
\setlength{\tabcolsep}{2.2pt}
\caption{Post-study replay of unauthorized evaluator drift.}
\label{tab:admission_objects}
\begin{tabularx}{\columnwidth}{@{}>{\raggedright\arraybackslash}p{0.39\columnwidth} >{\raggedright\arraybackslash}X@{}}
\toprule
Validation layer & Decision on two evaluation artifacts \\
\midrule
Run completion & both return terminal metrics \\
Per-run lineage & each metric is traceable to its evaluator revision \\
Per-run artifact validation & both artifacts are internally complete \\
Post-study pair admission & reject: evaluator mismatch is outside the approved contrast \\
\bottomrule
\end{tabularx}
\end{table}

\subsection{Pivot-conditioned video discovery}
The Deep Structured Semantic Model (DSSM) introduced learned semantic matching~\cite{huang2013learning}. Two-tower recommenders~\cite{covington2016deep} and large-scale search retrievers~\cite{huang2020embedding,liu2021que2search} scale this pattern. Session models condition on a current item or interaction history~\cite{hidasi2016session,kang2018self}, while personalized search adds user context to a typed query~\cite{teevan2005personalizing,bennett2012modeling}. VDD combines these settings. A request contains both a seed video and a machine-generated pivot query, while the system retains search relevance and a multi-stage retrieval funnel. Its interaction head re-scores approximate nearest-neighbor (ANN) candidates together with request features. Our contribution operates at the experiment-control layer above this model architecture.

\subsection{GPU retrieval infrastructure}
Scalable ANN systems~\cite{malkov2018efficient,subramanya2019diskann,chen2021spann,johnson2019billion} and GPU retrieval infrastructure~\cite{faiss_cuvs_2025,silvertorch} provide the serving substrate for our study. EvoPilot addresses experiment control across model development, publication, and evaluation rather than ANN kernel or serving design.

\section{Problem Formulation and System Overview}
\label{sec:overview}

\subsection{Pivot-conditioned video discovery}
VDD is a follow-on discovery experience entered from a video that the user has opened. It presents another set of videos for deeper exploration. Unlike conventional search, the user need not formulate a new query: the system derives a pivot query from the opened video and session context.

We formalize a VDD request as a triple $(s, q, u)$: the opened \emph{seed} video $s$, the machine-generated \emph{pivot query} $q$, and a user/context profile $u$. From a servable inventory $\mathcal{D}$, the system must retrieve a short ranked list of videos worth watching next. Retrieval is conditioned jointly on $(s, q, u)$. This differs from intent-driven search, which uses a user query alone, and from next-item recommendation, which uses $s$ and $u$ without a query, search relevance, or a retrieval funnel. The desired score is a learned function $f(s,q,u,d)$.

\subsection{Model-based retrieval pipeline}
\label{sec:mbr}
{\emergencystretch=.5em
MBR is a learned dense candidate generator, not the downstream ranker that orders an already retrieved list. The pipeline approximates $f$ in two stages. It first performs approximate nearest-neighbor (ANN) retrieval using dot products. When enabled, it then applies non-linear candidate-conditioned scoring. The system must perform these computations over candidates drawn from hundreds of millions of videos while meeting online throughput and latency constraints. The two-tower first stage maps the request and every video into a shared space,
\par}
\begin{equation}
 h_r=Q(s,q,u), \qquad h_d=D(d), \qquad a(d)=\langle h_r,h_d\rangle .
\end{equation}
The request tower $Q$ consumes the pivot query, user context, and request-side features of the seed video. The candidate tower $D$ consumes only candidate-video features. Both outputs are normalized, so dot product and cosine similarity induce the same ordering. This separation is operationally important: $D$ is evaluated offline over the servable inventory, while $Q$ runs once per request.

Training pairs each request $i$ with an engaged video $d_i$. The primary two-tower objective is out-of-session in-batch InfoNCE,
\begin{equation}
 \mathcal{L}_{\mathrm{NCE}}=-\frac{1}{B}\sum_{i=1}^{B}
 \log\frac{\exp(a_i(d_i)/\tau)}
 {\sum_{j\in\mathcal{B}_i}\exp(a_i(d_j)/\tau)},
\end{equation}
where $\mathcal{B}_i$ excludes other examples from the same session so that plausible co-engagements are not treated as negatives. An existing non-linear interaction head is also instantiated during training. It evaluates a candidate-specific function $G(h_r,h_d,a(d))$ and supplies weighted binary-cross-entropy losses for click and engagement-derived auxiliary labels. A relevance output is retained for interface compatibility but has zero training weight in the studied VDD configuration. The total objective is the weighted sum of InfoNCE and these auxiliary losses; consequently, the interaction head can shape the shared representation even when its scores are not used at inference.

Publishing applies $D$ to the servable inventory and builds an ANN index of hundreds of millions of videos on an existing GPU platform~\cite{silvertorch}. Candidate embeddings and the index are refreshed hourly, but the neural weights are not retrained every hour. At request time, ANN retrieves the $P$ candidates with the largest affinity $a(d)$. When candidate-conditioned scoring is enabled, the interaction head evaluates those $P$ candidates. A deterministic weighted combination of task scores then retains the final top $K$ for the downstream retrieval funnel. Thus $P$ controls how many candidates enter candidate-conditioned scoring, whereas $K$ is the MBR output depth.

Training jointly produces the two towers and interaction head as one checkpoint. Publication transforms that checkpoint into a serving artifact containing the request-side graph and candidate index. Offline replay uses a larger evaluation artifact derived from the same checkpoint. It reconstructs a candidate pool and measures whether held-out engaged videos survive the top-$K$ output (Section~\ref{sec:metrics}). These artifact boundaries allow the workflow to resume from a checkpoint or evaluation artifact that has already satisfied its post-conditions, avoiding unnecessary retraining or republication.

Before EvoPilot, manual experiments had \textbf{not} established a consistent improvement from interaction-head scoring. Experts therefore did not advance the direction, and the baseline left the head disabled. With $P=K=3000$, this baseline returned the two-tower ANN order. The treatment activates the already-trained interaction head at publication and serving time; it does not change training. As Section~\ref{sec:rq2} shows, a subsequent primitive autoresearch attempt incorrectly reinforced the negative decision. EvoPilot was then introduced, and its required review traced the apparent regression to a mismatch in effective output depth. The resulting matched comparison and isolated online A/B test held $K$ and the trained checkpoint fixed, and the only change was whether interaction-head scores were used. Because the head is the sole consumer of the $P$ retrieved candidates, an arm with the head disabled returns the two-tower order and is invariant to $P$.

An MBR experiment therefore crosses source and model lineage, training data, candidate-index publishing, serving configuration, and replay evaluation. A change may alter $Q$, $D$, $G$, the task-score weights, $P$, or $K$. A comparison between two completed workflows is valid only when every undeclared dimension remains aligned. EvoPilot can drive experiments on any of these model or configuration dimensions. Its contribution is to control and verify execution across this multi-system path, not to introduce a new VDD model architecture.

\section{EvoPilot: Protocol-Driven Agentic Experimentation}
\label{sec:evopilot}

\textsc{EvoPilot} controls the multi-stage VDD experiment workflow in Section~\ref{sec:overview} by separating execution guidance from evidence admission. A versioned MBR procedural skill and an execution adapter translate approved experiment specifications into operations. A human-approved protocol defines the intended comparisons, quantities extracted from terminal artifacts support deterministic verification, and human review completes evidence admission. The evidence unit is therefore an admitted control--treatment comparison, not a completed run. Agent proposals, configuration settings, skill invocations, and self-reports are inputs to this process; none alone is execution authority or scientific evidence.

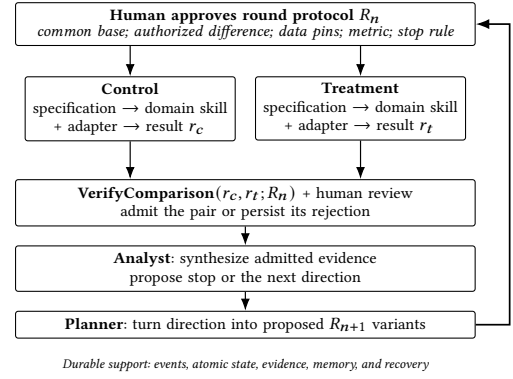
\begin{figure}[t]
\centering
\begin{tikzpicture}[
  font=\scriptsize,
  box/.style={draw, rounded corners=1pt, align=center, inner sep=2.2pt},
  arm/.style={box, text width=0.31\columnwidth},
  wide/.style={box, text width=0.70\columnwidth},
  flow/.style={-{Latex[length=1.6mm]}, line width=0.45pt},
  loop/.style={-{Latex[length=1.8mm]}, line width=0.65pt}
]
\node[wide] (approve) {\textbf{Human approves round protocol $R_n$}\\[-1pt]
  \textit{common base; authorized difference; data pins; metric; stop rule}};
\node[arm, below=4mm of approve, xshift=-0.18\columnwidth] (control)
  {\textbf{Control}\\specification $\rightarrow$ domain skill\\+ adapter $\rightarrow$ result $r_c$};
\node[arm, below=4mm of approve, xshift=0.18\columnwidth] (treatment)
  {\textbf{Treatment}\\specification $\rightarrow$ domain skill\\+ adapter $\rightarrow$ result $r_t$};
\node[wide, below=5mm of control, xshift=0.18\columnwidth] (verify)
  {\textbf{VerifyComparison}$(r_c,r_t;R_n)$ + human review\\admit the pair or persist its rejection};
\node[wide, below=2.5mm of verify] (analyst)
  {\textbf{Analyst}: synthesize admitted evidence\\propose stop or the next direction};
\node[wide, below=2.5mm of analyst] (planner)
  {\textbf{Planner}: turn direction into proposed $R_{n+1}$ variants};

\draw[flow] ([xshift=-0.18\columnwidth]approve.south) -- (control.north);
\draw[flow] ([xshift=0.18\columnwidth]approve.south) -- (treatment.north);
\draw[flow] (control.south) -- ([xshift=-0.18\columnwidth]verify.north);
\draw[flow] (treatment.south) -- ([xshift=0.18\columnwidth]verify.north);
\draw[flow] (verify) -- (analyst);
\draw[flow] (analyst) -- (planner);
\coordinate (loopbottom) at ([xshift=4.5mm]planner.east);
\coordinate (looptop) at (loopbottom |- approve.east);
\draw[loop] (planner.east) -- (loopbottom) -- (looptop) -- (approve.east);

\node[align=center, font=\tiny\itshape, below=1.5mm of planner]
  {Durable support: events, atomic state, evidence, memory, and recovery};
\end{tikzpicture}
\caption{Comparison-centric, cross-round EvoPilot loop. One authorized contrast is shown; multi-variant rounds verify each separately. Admitted evidence shapes the next proposal, which returns to human approval.}
\Description{A human-approved protocol creates control and treatment specifications that execute in isolation through a domain skill and adapter. Artifact-backed results undergo pair-level verification and human review. An analyst synthesizes admitted evidence into a next-direction recommendation, and a planner proposes next-round variants that loop back to human approval. Durable state connects rounds.}
\label{fig:method}
\end{figure}

Figure~\ref{fig:method} shows the resulting loop. Agent sessions, remote workflows, and research rounds can have different lifetimes, so durable state connects them. Humans retain control of research direction, approval, metric semantics, exceptions, and stopping. The system gained new controls as failures were observed. This section describes the system at manuscript cutoff. Section~\ref{sec:hazards} distinguishes controls used during the study from post-study controls evaluated only by replay.

\subsection{Domain skill as the execution foundation}
\label{sec:adapter}
Workers do not infer MBR procedures from research prompts. The versioned domain skill defines configuration discovery, local small-scale preflight, the train $\rightarrow$ publish-for-evaluation $\rightarrow$ evaluate DAG, monitoring, partial recovery from valid upstream artifacts, and metric extraction. For each approved variant, the execution adapter maps its typed specification to these operations. It also declares workflow and artifact dependencies, allowed tool classes, and required post-conditions. The skill explains how to operate the MBR stack, while the adapter binds one approved variant to those operations. Neither can admit evidence. \textsc{VerifyComparison} establishes mechanical eligibility, and human review completes admission.

\subsection{Comparison protocol and admission}
\label{sec:comparison_protocol}
For EvoPilot, a human-approved round protocol $R$ names the common base, data pins, required terminal artifacts, metric, closure rule, and one or more authorized contrasts. Each contrast identifies a control arm, a treatment arm, and the difference allowed between them. Once both arms reach a terminal state, the fail-closed verifier $\operatorname{Verify}_R(c,t)$ checks the pair. Both arms must have succeeded. Artifact-extracted quantities must match the approved specifications, the realized difference must equal the authorized treatment, and stages unaffected by that treatment must have equivalent artifacts. Missing required evidence causes rejection. Admission is
\begin{equation}
\label{eq:admission}
\mathcal{A}(c,t;R)=
\mathbb{1}[\operatorname{Verify}_R(c,t)]
\cdot\mathbb{1}[\operatorname{Review}_R(c,t)].
\end{equation}
$\operatorname{Review}_R$ records human judgment about metric semantics and exceptions. Trackers may record each arm separately, but \textsc{VerifyComparison} joins their attestations at the pair level. A multi-variant round is evaluated one contrast at a time. Completion of other variants cannot make an invalid pair admissible.

The historical primitive-autoresearch attempt illustrates this distinction. Both arms produced terminal metrics and declared $P{=}3000$ and $K{=}600$. Both realized $P_{\mathrm{eff}}{=}3000$, but the control realized $K_{\mathrm{eff}}{=}3000$ while the treatment realized $K_{\mathrm{eff}}{=}600$. After EvoPilot was introduced, its required human-gated review rejected the model attribution and triggered an artifact-level investigation. The automatic comparator between declared and effective funnel settings was added only during post-study hardening and is evaluated by replay. A separate evaluator-drift incident showed the same pair-level requirement: two complete artifacts were still incomparable because they came from different evaluator revisions. Admission therefore depends on the realized execution of both arms, not on whether each run completed. Section~\ref{sec:protocol} gives a concrete protocol and attestation interface.

\subsection{Round protocol}
\label{sec:loop_arch}
A \emph{research direction} records the hypothesis tree, baseline, allowed treatment dimensions, primary metric, and stopping rule. At round $n$, a planner reads this direction protocol together with prior plans, results, takeaways, and human edits. It then proposes variants and contrasts mapped to nodes in the hypothesis tree. Prior evidence can close a branch, motivate a new branch, or focus the next round on a remaining uncertainty. The researcher may edit, reject, or approve the plan. Approval applies only to the current round; approval for round $n$ cannot authorize round $n{+}1$.

After approval, a base builder materializes the shared source changes, data pins, and configuration as one round base. Each variant inherits that base and adds only its declared treatment. The variant specification names the base, treatment, data pins, stages, expected direction, and model roles. The protocol treats this specification as read-only, although it is not cryptographically immutable.

Artifact reuse also acts as a validity control. For a fixed checkpoint, the system computes an equivalence key from the checkpoint identifier, code changes, pinned data, and publication settings. Audited fields that affect only evaluation are excluded. One publication produces the evaluation artifact, and equivalent variants reuse it; changes to training configuration or model lineage prohibit reuse. The three-arm candidate-depth sweep therefore shared one artifact. This avoided two publications and the variation they could introduce, while leaving candidate depth as the only realized difference.

Missing terminal results block closure by default. A human may explicitly close a partial round, but the exception is logged and comparisons involving failed variants remain inadmissible. At closure, a results analyst maps admitted outcomes to the hypothesis tree. It records invalid measurements and operational failures separately, then recommends either stopping or pursuing a high-level next direction. If the research continues, the planner combines that recommendation with the cross-round record to propose new hypotheses and variants. This handoff from evidence synthesis to a new proposal is the autoresearch step. The new plan still requires fresh human approval.

Algorithm~\ref{alg:loop} gives the complete protocol. Its assertions cover violations that would change the interpretation of a comparison. Hypothesis quality and unexpected scientific judgments remain outside this mechanical boundary.

\begin{algorithm}[t]
\scriptsize
\setlength{\itemsep}{0pt}
\caption{Human-gated experiment round}
\label{alg:loop}
\begin{algorithmic}[1]
\STATE \textbf{input:} research direction $H$, evidence ledger $D$, lesson memory $M$
\FOR{$n=1$ \TO round cap}
  \STATE $C\leftarrow\textsc{Retrieve}(H,D,M)$
  \STATE $\widehat R_n\leftarrow\textsc{ProposeRound}(H,C)$; $R_n\leftarrow\textsc{HumanApprove}(\widehat R_n)$
  \STATE $b\leftarrow\textsc{MaterializeBase}(R_n)$; \textbf{assert} \textsc{VerifyBase}$(b,R_n)$
  \FORALL{$v\in\textsc{Variants}(R_n)$ (concurrency capped)}
    \STATE $c\leftarrow\textsc{LeaseIsolatedClone}()$; materialize $\mathit{spec}_v$
    \STATE $r_v\leftarrow\textsc{ExecuteSkillAdapter}(c,b,\mathit{spec}_v)$
    \STATE append state; mark $r_v$ invalid if verification fails
  \ENDFOR
  \FORALL{$(c,t)\in\textsc{Contrasts}(R_n)$}
    \STATE $a_{ct}\leftarrow\textsc{VerifyComparison}(r_c,r_t,R_n)\land\operatorname{Review}_{R_n}(c,t)$
    \STATE admit $(c,t)$ only if $a_{ct}$
  \ENDFOR
  \STATE $s_n\leftarrow\textsc{Analyze}(H,R_n,\{r_v\},\{a_{ct}\})$; persist hypothesis update and next/stop
  \STATE update $D/M$/guards; human reviews exceptions; \textbf{break} if $s_n$ stops
\ENDFOR
\end{algorithmic}
\end{algorithm}

\subsection{Agent topology and system boundary}
\label{sec:agents}
Six versioned role families implement the protocol through typed inputs, capabilities, and outputs. The \emph{round planner} proposes a plan for approval, and the \emph{base builder} emits the shared change set. Each resumable \emph{experiment worker} receives one approved specification and operates in an isolated environment through the domain skill and adapter. It emits a terminal result and execution trace. The read-only \emph{results analyst} uses terminal and admission records to produce a hypothesis update, round summary, and recommendation to stop or continue in a new direction.

Separate roles handle operational failures. The read-only \emph{periodic monitor} classifies each unseen event as ignorable, repairable, or requiring escalation. The \emph{failure handler} first performs a read-only diagnosis. It can emit an operational fix only through a separately gated repair invocation. Table~\ref{tab:agents} summarizes these protocols.

\begin{table}[t]
\centering
\scriptsize
\setlength{\tabcolsep}{2.2pt}
\caption{Agent roles, capabilities, and typed outputs.}
\label{tab:agents}
\begin{tabularx}{\columnwidth}{@{}>{\raggedright\arraybackslash}p{0.25\columnwidth} >{\raggedright\arraybackslash}p{0.25\columnwidth} >{\raggedright\arraybackslash}X@{}}
\toprule
Role & Capability & Typed output \\
\midrule
Round planner & read-only evidence & variant plan for fresh approval \\
Base builder & edit shared baseline & source/configuration change set \\
Experiment worker & isolated skill/adapter execution & result record + trace \\
Results analyst & read admitted results & hypothesis update; stop/next direction \\
Failure triage/repair & diagnose; gated repair & fault verdict / operational fix \\
Periodic monitor & read incremental events & ignore, repair, or escalate \\
\bottomrule
\end{tabularx}
\end{table}

The orchestrator and verifier are deterministic programs, not agents. Control code retrieves state, launches roles, gates rounds, and persists events. \textsc{VerifyComparison} checks artifact-backed pairs against their protocols. Humans retain authority over research direction, metric semantics, approvals, exceptions, and stopping. A common launcher applies versioned tool permissions and records the role, model alias, session, status, and duration of each invocation. Experiment workers receive a versioned MBR procedural skill in addition to their role protocol and adapter access. Planning, experiment modification, result synthesis, and repair used Claude Opus 4.7/4.8; bounded read-only monitoring used Claude Haiku.

\subsection{Persistent evidence and enforcement}
\label{sec:memory}
\label{sec:verifiers}
\label{sec:recovery}
Durable support has three scopes (Appendix Table~\ref{tab:memory}). The run ledger stores typed events and atomic variant and round results, allowing workflows to resume and supporting regression checks. Before planning, structured lookup retrieves prior plans, results, takeaways, and experiment summaries from program memory. The incident catalog contains 38 searchable records in six categories. Its text remains advisory until a lesson is encoded as an executable check. A cursor-based monitor reads only unseen events and chooses whether to ignore, repair, or escalate them without overwriting the raw evidence.

Each stage persists and synchronizes its state before recording completion, so a crash replays an uncommitted stage rather than silently skipping it. Workflow/model identifiers and round-specific approval make recovery inspectable. In one study round, execution resumed from an unfinished base-construction stage after storage loss while retaining approval and prior evidence.

Enforcement relies on artifacts rather than LLM self-assessment. Deterministic checks cover base ancestry, effective overrides, data drift, and local validation. On an unrecoverable failure, the controller records the evidence and either stops execution or escalates to a human. Pair-level verification implements the protocol in Section~\ref{sec:comparison_protocol}. Both arms must succeed and descend from the approved base, their data pins must match, and their realized difference must equal the authorized treatment. Stages unaffected by the treatment must also have equivalent artifacts. A configured setting proves intent, not execution. Post-study hardening added a comparator over terminal attestations that requires $(P_{\mathrm{decl}},K_{\mathrm{decl}})$, $(P_{\mathrm{eff}},K_{\mathrm{eff}})$, evaluator identity, and provenance. The verifier rejects missing attestations, self-reported numbers without artifact provenance, and unauthorized differences. A human may allow a workflow to continue after an exception, but its comparison remains explicitly non-admitted. Table~\ref{tab:invariants} reports when each control was introduced.

\subsection{Control evolution from observed failures}
\label{sec:guards}
Each reviewed incident identifies the violated invariant and, when feasible, leads to a preflight, post-condition, wrapper, assertion, or gate. We do not credit a control with preventing the incident that motivated it. Section~\ref{sec:hazards} therefore reports failures detected during the longitudinal study separately from mutation tests of the final controls. This separation makes the system's evolution auditable and avoids presenting later automation as if it had operated earlier.

\begin{table}[t]
\centering
\scriptsize
\setlength{\tabcolsep}{2.2pt}
\caption{Control coverage at manuscript cutoff. Executable controls block normal progression on failure.}
\label{tab:invariants}
\begin{tabularx}{\columnwidth}{@{}>{\raggedright\arraybackslash}p{0.22\columnwidth} >{\raggedright\arraybackslash}p{0.45\columnwidth} >{\raggedright\arraybackslash}X@{}}
\toprule
Invariant & Control & Introduction \\
\midrule
Round authority & approval tied to the round identifier & outset \\
Environment & dedicated clone or fail & outset \\
Change identity & lineage, nonempty code diff, effective-configuration readback & strengthened \\
Data identity & date pins, partition and drift checks & strengthened \\
Train/eval order & configuration and artifact-reuse checks & after leakage \\
Completeness & failure gate; explicit partial closure & during study \\
Metric context & declared/effective $(P,K)$, evaluator ID, provenance; missing/mismatch fails closed & post-study \\
\bottomrule
\end{tabularx}
\end{table}

\subsection{Concrete comparison protocol}
\label{sec:protocol}
The following instance expresses a serving treatment over two authorized fields: enabling the interaction head also authorizes the candidate depth $P$ that the head rescores. Digests stand for versioned identifiers rather than concrete implementation paths.

\begin{lstlisting}[basicstyle=\ttfamily\scriptsize,frame=single]
contrast: head_on_vs_off
allowed_diff: [serve.interaction_head, serve.P]
control:
  base: B7; data: {train: T4, pool: P9, eval: E3}
  serve: {interaction_head: false, P: 3000, K: 3000}
treatment:
  base: B7; data: {train: T4, pool: P9, eval: E3}
  serve: {interaction_head: true,  P: 6000, K: 3000}
both:
  stages: [train, publish, evaluate]
  metric: {name: click_hit_rate, depth: 3000}
  require: [terminal, lineage, realized_override,
            candidate_count_attestation, evaluator_id]
procedural: [metric_semantics]
\end{lstlisting}

The arm blocks and \texttt{allowed\_diff} form the protocol used by $\operatorname{Verify}_R$ in Equation~\ref{eq:admission}. Each result adds terminal status, realized overrides, source and model lineage, workflow and artifact identifiers, metrics, recovery history, declared and effective $(P,K)$, and evaluator identity with provenance. The verifier rejects missing evidence, checks artifact-derived quantities against the corresponding arm, and requires that only the listed field paths differ. Stages unaffected by the serving-only treatment must share equivalence keys; treatment-affected artifacts must descend from the realized specification. $\operatorname{Review}_R$ records human adjudication of metric semantics. Rejected predicates are persisted, and their measurements cannot enter the evidence ledger.

Further operational details are reported in Appendix~\ref{sec:implementationapp}.

\section{Empirical Study}
\label{sec:eval}

We evaluate \textsc{EvoPilot} in a 37-day longitudinal VDD campaign across seven research directions. Expert-led manual experiments predated this campaign. The campaign opened with a primitive autoresearch attempt, and EvoPilot was introduced only after that attempt produced the invalid interaction-head conclusion analyzed in RQ2. VDD model developers then used EvoPilot in an evaluation connected to training, publishing, and evaluation systems, not to an offline simulator. A comparison enters the evidence ledger only when its specification, execution record, and terminal artifacts show that both arms realized the approved common base and only the authorized difference. Manual sweeps provide context but are not pooled with admitted comparisons.

The primary evaluation target is the experiment-control method. Paired replays test the admission guards. Execution records measure terminal coverage, recovery, and artifact reuse, while the longitudinal case documents the resulting research decision. The online A/B is a separate downstream endpoint. It causally estimates only the selected \emph{model treatment} and is not evidence for EvoPilot's effectiveness.

\subsection{Protocol, metrics, and auditable cohort}
\label{sec:metrics}
The agent setup follows Sections~\ref{sec:agents}--\ref{sec:memory}. Scientific and repair roles used Claude Opus 4.7 and 4.8. Claude Haiku handled bounded read-only monitoring under the same typed role protocols. Each run retained its role definition, rendered prompt or prompt hash, model alias, and session identifier. Worker concurrency was typically three and never exceeded five. A local preflight took about 40 minutes, and a successful remote workflow usually took 1--3 hours. Workers also supported longer-running tasks and partial recovery. Every round and every research-level exception required human approval.

Offline evaluation replays held-out pairs against one candidate pool. Its primary selection metric is click hit-rate at depth $K$,
\begin{equation}
H_{\mathrm{click}}@K =
\frac{\sum_i \mathbb{1}[y_i^{\mathrm{click}}=1 \land d_i\in\mathrm{top}K_i]}
     {\sum_i \mathbb{1}[y_i^{\mathrm{click}}=1]}.
\end{equation}
Other label heads and candidate-pool statistics are diagnostic. Two readings are comparable only when their data dates, $K$, candidate depth, search parameters, and evaluator revision align as required by the protocol. Online serving uses $K{=}3000$; we report $K{=}600$ only as a diagnostic cut. Offline readings are descriptive point estimates, and we do not report inferential confidence intervals for them. The $\pm0.36$ percentage-point band used below is an empirical tolerance from repeat publication. Two publications with the same model, configuration, and pinned data differed by 0.36 points because publish-time shuffling produced slightly different candidate sub-pools. This band measures operational variation, not sampling uncertainty.

The online endpoint belongs to the online metric family \emph{Good Search Result Rate for Retention} (GSRR). Its VDD slice is a session-level rate. For pivot sessions $j\in\mathcal{S}$, let $E_j$ indicate VDD metric eligibility and $G_j$ indicate that the session contains the qualifying engagement defined by the metric. Then
\begin{equation}
\operatorname{GSRR}=\frac{\sum_{j\in\mathcal{S}} E_jG_j}{\sum_{j\in\mathcal{S}}E_j}.
\label{eq:gsrr}
\end{equation}
Thus VDD GSRR is the fraction of eligible pivot sessions with qualifying engagement. We report relative lift rather than a percentage-point difference,
\[
100\times\frac{\operatorname{GSRR}_t-\operatorname{GSRR}_c}
{\operatorname{GSRR}_c}\%.
\]
Offline hit-rate and online GSRR have different populations and denominators.

Agent actions are taken from retained role outputs and result records. Human actions include both approval and scientific adjudication. The seven directions share infrastructure and model lineage, so they are not independent replicates.

We distinguish three outcomes before analyzing a research direction. A \emph{scientific result}, including a non-improving result, requires a realized treatment, an admissible comparison, and a terminal metric. An \emph{invalid measurement} produces a number but violates the experiment protocol. It can motivate a new guard, but it cannot be used to rank models. An \emph{operational failure}, such as an out-of-memory training run or a missing data partition, has no model estimand. This classification prevents an execution failure from being reported as negative evidence about a model.

The 37-day campaign inventory contains seven directions, 20 human-approved rounds, and 68 configurations with reported metrics. It covers both the primitive attempt and the subsequent EvoPilot period. Because older records omit some launches and terminal failures, the 68 configurations cannot serve as the denominator for workflow-attempt rates. We used an outcome-independent completeness screen for operational statistics. An included round had to retain its approval and plan, every launched variant and terminal state, the logical train/publish/evaluate stages, retry and repair history, and elapsed time. Nine EvoPilot rounds met every requirement: one score round, one depth round, and seven capacity or loss rounds. The excluded set contains three rounds on the interaction head, one on another task score, two on candidate depth, two on architecture replay, one on the evaluator, and two on seed features. At least one required operation field in each excluded round was not recorded under the stable schema. These rounds remain part of the scientific context and incident analysis, but not of the EvoPilot operation-rate denominators. Table~\ref{tab:methodoutcomes} summarizes the resulting inclusion flow and operation census, and Appendix Table~\ref{tab:operationaudit} lists the included rounds.

\begin{table}[t]
\centering
\scriptsize
\setlength{\tabcolsep}{2.4pt}
\caption{Inclusion flow and complete operation census for the auditable cohort.}
\label{tab:methodoutcomes}
\begin{tabularx}{\columnwidth}{@{}>{\raggedright\arraybackslash}p{0.27\columnwidth} >{\raggedright\arraybackslash}X@{}}
\toprule
Operation & Observed record \\
\midrule
Round inclusion & 9/20 rounds included; 11 excluded for incomplete operation fields \\
Workflow closure & 29/29 retained terminal records: 27 scientific results; two OOM failures \\
Stage construction & 22 train attempts (20 succeeded); 22 publishes; 27 evaluations; 71 total \\
Primary role assignments & 56: 9 planner, 9 base-builder, 29 worker, 9 analyst \\
Elapsed operation & $\sim$97 h summed orchestrator wall-clock across the nine rounds \\
Repair and recovery & 24 agent retries after local fixes; 6 operator repairs; 6 subworkflows and one round recovered; one escalation \\
Artifact reuse & two artifacts served seven evals; five publications avoided; $\sim$5 GPU-hours documented for depth \\
\bottomrule
\end{tabularx}
\end{table}

The stage counts do not represent 29 repetitions of one fixed pipeline, because variants ran only the stages they needed. The 22 capacity or loss workflows attempted training. Two stopped after an out-of-memory failure, while the other 20 continued to publication and evaluation. The score and depth rounds were evaluation-only sweeps in which seven evaluations reused two shared publications. The cohort therefore contains 22 training stages, 22 publication stages, and 27 evaluation stages, for 71 logical stages in total. Within this cohort, 29/29 measures terminal-record coverage and 27/29 measures completion with a scientific result. Neither number is a success rate for all 20 rounds.

\subsection{RQ1: Can controls distinguish clean states from known faults?}
\label{sec:rq1}
For each invariant, we paired an admissible artifact or state with a named fault and invoked the deterministic guard without LLM judgment. Seven pairs cover workflow controls. Post-study hardening added three identity pairs: repaired versus historical effective funnel, authorized versus undeclared $P$ difference, and unchanged versus changed evaluator identity. All ten clean fixtures were admitted, and all ten corresponding faults were blocked (Table~\ref{tab:faultreplay}). The verifier also rejected terminal readings that were missing artifact-backed provenance or were only self-reported.

\begin{table}[t]
\centering
\scriptsize
\setlength{\tabcolsep}{2.5pt}
\caption{Paired replay/mutation: 10/10 clean fixtures admitted and 10/10 corresponding faults blocked.}
\label{tab:faultreplay}
\begin{tabularx}{\columnwidth}{@{}>{\raggedright\arraybackslash}p{0.25\columnwidth} >{\raggedright\arraybackslash}p{0.30\columnwidth} >{\raggedright\arraybackslash}X@{}}
\toprule
Invariant & Clean fixture & One-field mutation \\
\midrule
Round authority & current round proceeds & stale approval rejected \\
Change identity & nonempty code diff passes & absent code diff rejected \\
Override identity & effective configuration matches & wrong setting rejected \\
Data identity & pinned dates match & one date drift rejected \\
Artifact reuse & equivalent key passes & incompatible reuse rejected \\
Train/eval order & eval follows train & leakage rejected \\
Completeness & all arms terminal & one failed arm blocks closure \\
Effective funnel & both arms realize $(3000,600)$ & control realizes $(3000,3000)$ \\
$P$ authorization & declared depth treatment & undeclared $P$ difference rejected \\
Evaluator identity & same revision & cross-revision pair rejected \\
\bottomrule
\end{tabularx}
\end{table}

\paragraph{Control evolution from observed faults.}
\label{sec:hazards}
The replay above evaluates the controls as they existed at manuscript cutoff; not every control operated prospectively. Four observed failures led to new controls: funnel mismatch to fail-closed $K/P$ admission, evaluator drift to evaluator-identity checks, train/eval leakage to a date-order guard, and lineage or no-op edits to change, lineage, and readback checks. Funnel and evaluator identity became executable only during post-study hardening. Evaluator semantics beyond the recorded software version still require human review.

The funnel incident and its repair are analyzed in RQ2. We reproduced the evaluator failure by running one published model with evaluator revisions from before, during, and after a metric-aggregation change. Readings at each $K$ remained stable, but the aggregate metric changed meaning. In the leakage case, a favorable result used an evaluation date inside the candidate model's training window, so the result was retracted. In the lineage case, history rewriting removed the intended edit but left a plausible revision chain. A separate train/serve coverage analysis found a seed identifier in nearly all training examples but absent online. Offline feature coverage alone therefore cannot establish train/serve consistency.

At study conclusion, the incident catalog contained 38 records in six categories. The post-study comparator now writes one record for every in-round contrast, which provides a denominator for future admission rates. It cannot reconstruct the missing denominator for the earlier study period.

\subsection{RQ2: Can EvoPilot overturn an invalid primitive-autoresearch conclusion?}
\label{sec:rq2}
Before EvoPilot, expert-led exploratory work had examined larger candidate towers, additional query and seed representations, alternative task combinations, and interaction-head inference without finding a stable improvement. For the interaction head specifically, exploratory comparisons were mixed: at $K{=}3000$ the two arms were within a few tenths of a percentage point of each other, while at $K{=}600$ they differed by several percentage points in the opposite direction. The $K{=}600$ arms also differed in candidate depth $P$. Experts therefore did not advance the direction, although the hypothesis remained scientifically unresolved.

The primitive autoresearch attempt revisited the direction and reported that hit rate fell by 22 percentage points. It attributed the decline to the interaction head, even though both arms declared $(P,K){=}(3000,600)$. On its face, this result reinforced the earlier decision not to proceed. We then introduced EvoPilot. Its required human-gated verification rejected the conclusion after finding that both arms realized $P_{\mathrm{eff}}{=}3000$, but the control realized $K_{\mathrm{eff}}{=}3000$ while the treatment realized $K_{\mathrm{eff}}{=}600$.

An extreme-$K$ falsification probe confirmed the mismatch. Before repair, changing $K_{\mathrm{decl}}$ from 600 to 1 barely changed the two-tower-only hit rate, which moved by about half a percentage point. This showed that $K_{\mathrm{eff}}$ remained 3000. The repair added top-$K$ truncation when candidates enter the evaluation pipeline and a second truncation before metric computation. Afterward, the same probe produced the expected monotone collapse: the $K{=}600$ reading fell far below its pre-repair value, and the $K{=}1$ reading was near zero, confirming that the repair took effect. The first EvoPilot version relied on the human gate. The automatic identity verifier was added after this incident, so we evaluate it by replay rather than claim prospective detection.

After repair, a re-test with both arms realizing $(3000,600)$ measured an increase of $4.8$ percentage points. This re-test also changed the base model, so it estimates a bundled configuration effect. A later matched-lineage comparison fixed $K{=}3000$ and the trained checkpoint, running the control at $P{=}3000$ with the head off and the treatment at $P{=}6000$ with the head on, and measured an increase of $3.20$ percentage points. Because the control disables the head, its top-$K$ output is the two-tower affinity order and is invariant to $P$, this difference isolates the interaction head at its selected operating point. Table~\ref{tab:interaction_evidence} distinguishes these estimands. EvoPilot thus replaced a falsely negative result with an admissible estimate of the head-only effect and advanced the treatment.

\begin{table}[t]
\centering
\scriptsize
\setlength{\tabcolsep}{2.4pt}
\caption{Interaction-head comparisons across funnel configurations. Offline replay estimates are descriptive.}
\label{tab:interaction_evidence}
\begin{tabularx}{\columnwidth}{@{}>{\raggedright\arraybackslash}p{0.23\columnwidth} >{\raggedright\arraybackslash}X r@{}}
\toprule
Stage & Estimand / comparability & Result \\
\midrule
Expert-led manual evidence & mixed estimands; one pair coupled & direction not advanced \\
Primitive autoresearch baseline & mismatched $K$; incorrect attribution & $-22$\,pp \\
Diagnostic re-test & bundled effect; $(P,K){=}(3000,600)$ & $+4.8$\,pp \\
Matched-lineage ablation & head-only effect; $K{=}3000$, $P{:}\,3000\!\rightarrow\!6000$ & $+3.20$\,pp \\
Online A/B & head-only inference treatment; fixed remainder & $+0.66\%$ rel. \\
\bottomrule
\end{tabularx}
\end{table}

\subsection{RQ3: How do admitted outcomes constrain subsequent research?}
\label{sec:nulls}
RQ2 shows how an admitted improvement advanced a treatment. Iteration must also decide which directions to stop or redirect. Artifact reuse held publication and candidate-pool coverage fixed within the score and depth sweeps. Removing an unsupervised score term changed click $H@3000$ by $-0.17$ points, within the $\pm0.36$-point operational band, so the study selected the simpler click-only blend without claiming equivalence. Adding an engagement score reduced click hit rate by $1.53$ points. Results also declined beyond 6k candidates, and four in-session-negative weights changed click hit rate by only $[-0.36,+0.07]$ points while the nominal best reduced an auxiliary slice by $1.23$ points. These admitted results rejected the additive score, fixed depth at 6k, and stopped the loss direction. A no-op capacity patch is instead a change-identity failure (RQ1), not a model result.

\subsection{RQ4: What downstream outcome followed the admitted model decision?}
\label{sec:rq4}
We use the online A/B only to establish the product relevance of the selected model. The treatment was evaluated in a randomized online experiment that assigned user accounts to control and treatment in equal proportions, quantifying performance before the subsequent default configuration change. The account was both the randomization unit and the analysis unit, with no higher-level clustering. Control and treatment each contained on the order of tens of millions of assigned accounts, balanced to within a small fraction of a percent; because the outcome is the US and Canada VDD-pivot slice, only a small subset of treatment accounts contributed to the metric. The analysis covers a seven-day window and uses a regression-adjusted mean with outcomes from a separate seven-day pre-period. With the remainder of the system fixed, the estimated relative GSRR effect was $+0.66\%$, a statistically significant increase. Guardrail outcomes were not available in the retained dataset.

After the experiment review, the interaction-head treatment became the default VDD MBR online configuration and remained enabled at manuscript cutoff. The path uses the hourly refreshed index of hundreds of millions of videos described in Section~\ref{sec:mbr}. This is the outcome of the selected model change. It is separate from both the preceding 37-day agentic campaign and the randomized online experiment. Continued serving establishes the scope, while the randomized A/B remains the causal estimate of treatment effect. We infer no separate post-change effect.

\section{Discussion}
\label{sec:discussion}

\paragraph{The comparison is the unit of evidence.}
A completed run is not by itself scientific evidence. \textsc{EvoPilot} admits a comparison only when its protocol defines the contrast and terminal artifacts show the common base and authorized difference. Invalid measurements cannot rank models; admissible non-improvements can close a direction; operational failures have no model estimand. Humans define metric semantics and stopping rules, while the harness verifies execution.

\paragraph{Porting the method.}
Porting \textsc{EvoPilot} requires a versioned domain skill for procedure, a protocol mapper for the intended contrast, an execution adapter for workflow control, attestation extractors for realized artifact quantities, and an admission interface for verdicts. Agent proposals, memory, and self-reports remain untrusted; the trusted base is the approved protocol, artifact identities, deterministic extractors and verifiers, and durable verdict. Experts retain metric and stopping decisions. Any field that cannot be extracted deterministically from a trusted terminal artifact requires human review.

\paragraph{Limitations and threats to validity.}
Only nine rounds have complete operational records; invocation totals exist for two archived directories, so we do not extrapolate token cost or GPU occupancy. The primitive autoresearch baseline is a retained incident, not a concurrent randomized workflow comparison. Ten paired fixtures do not establish field recall or false-block rates, and the new identity gate has only replay evidence. Evaluator semantics still require human review. Offline confidence intervals, retained online guardrails, and cross-organization replication are unavailable.

\section{Conclusion}
\label{sec:conclusion}

We presented \textsc{EvoPilot}, a human-gated system that separates completed workflows from admissible comparisons through scoped roles, a versioned domain skill, durable evidence, and executable checks. The field study shows how failures can become pair-level controls: all ten clean replay fixtures were admitted and all ten faults blocked, including the historical funnel mismatch. Separately, the selected model produced a $+0.66\%$ relative GSRR.

\section*{Ethical Considerations}
The study followed applicable product-experiment review procedures, and we report only aggregate metrics. Creator-distribution effects were outside the endpoints and require separate monitoring.

\bibliographystyle{ACM-Reference-Format}
\bibliography{ref}

\appendix
\section{Operational details}
\label{sec:implementationapp}

\begin{table}[H]
\centering
\scriptsize
\setlength{\tabcolsep}{1.8pt}
\caption{Round-level operation census. R/V denotes rounds/variants, and T/P/E denotes logical train/publish/evaluate stages. A ``fix'' is an agent-initiated local correction; a ``repair'' is an operator action.}
\label{tab:operationaudit}
\begin{tabularx}{\columnwidth}{@{}>{\raggedright\arraybackslash}p{0.23\columnwidth} c c r >{\raggedright\arraybackslash}X@{}}
\toprule
Program & R/V & T/P/E & Hours & Retry / recovery / repair \\
\midrule
Score R1 & 1/4 & 0/1/4 & 2.3 & 1 fix; subworkflow recovered; one repair \\
Depth R3 & 1/3 & 0/1/3 & 2.2 & prelaunch repair \\
Capacity R1--R2 & 2/8 & 8/8/8 & 36.0 & 12 fixes; subworkflow recovered \\
Capacity R3 & 1/2 & 2/2/2 & 17.0 & 1 fix; subworkflow recovered; one repair \\
Capacity R4 & 1/2 & 2/2/2 & 8.0 & subworkflow recovered; one repair \\
Capacity R5 & 1/2 & 2/2/2 & 7.5 & 3 fixes; subworkflow recovered \\
Capacity R6 & 1/4 & 4/2/2 & 6.2 & 3 fixes; subworkflow recovered; one escalation \\
Loss-head R7 & 1/4 & 4/4/4 & 18.0 & 4 fixes; round recovered; two repairs \\
\midrule
Total & 9/29 & 22/22/27 & 97.2 & 24 fixes; 6 subworkflows + 1 round recovered; 6 repairs; 1 escalation \\
\bottomrule
\end{tabularx}
\end{table}

Selection depended on record completeness, not experimental outcome. Nine of the 20 rounds contained every census field, including the two arms that ended with OOM failures. The other 11 rounds remain outside the denominators for operation rates. Primary assignments comprise 9 planners, 9 base builders, 29 workers, and 9 analysts, for 56 assignments in total. Two archived raw ledgers contain 6,995 parseable events and 1,662 launches. Of those launches, 22 were primary-role invocations and 1,640 came from an older monitor that launched once per event. Later monitoring processed new events in batches, so we do not extrapolate the older invocation count.

\begin{table}[H]
\centering
\scriptsize
\setlength{\tabcolsep}{2.2pt}
\caption{Persistence scopes. Memory is advisory; enforcement requires executable checks.}
\label{tab:memory}
\begin{tabularx}{\columnwidth}{@{}>{\raggedright\arraybackslash}p{0.22\columnwidth} >{\raggedright\arraybackslash}p{0.35\columnwidth} >{\raggedright\arraybackslash}X@{}}
\toprule
Scope & Representation & Consumer / purpose \\
\midrule
Run ledger & events, atomic state, variant/round results & resume, traceability, regression check \\
Program memory & plans, prior results, relevant experiment summaries & next-round planner/workers \\
Incident catalog & 38 records in six categories & failure retrieval; guard design \\
\bottomrule
\end{tabularx}
\end{table}

\end{document}